\documentclass[preprint,showpacs,preprintnumbers,amsmath,amssymb]{revtex4}

\usepackage{subfigure}
\usepackage{graphicx}
\usepackage{dcolumn}
\usepackage{bm}

\begin{document}

\preprint{APS/123-QED}

\title{On Dynamic-order Fractional Dynamic System}

\author{HongGuang Sun$^{a,b}$}

\author{Hu Sheng$^{b,c}$}%
\author{YangQuan Chen$^{b}$}%
\author{Wen Chen$^{a,*}$}%
\author{ZhongBo Yu$^{d}$}%
\affiliation{%
a.Institute of Soft Matter Mechanics, Department of Engineering
Mechanics, Hohai University, No.1 XiKang Road, Nanjing, Jiangsu
210098, China\\
 b.Center for Self-Organizing and Intelligent Systems
(CSOIS), Department of Electrical and Computer Engineering, Utah
State University, 4160 Old Main Hill, Logan, UT 84322-4160, USA\\
c. Department of Electronic Engineering, Dalian University of
Technology, Dalian 116024, China\\
d. College of Hydrology and Water Resources, Hohai
University, No.1 XiKang Road, Nanjing, Jiangsu 210098, China\\
{*} Corresponding author: chenwen@hhu.edu.cn
}%


\date{\today}

\begin{abstract}
Multi-system interaction is an important and difficult problem in
physics. Motivated by the experimental result of an electronic
circuit element ``Fractor", we introduce the concept of
dynamic-order fractional dynamic system, in which the
differential-order of a fractional dynamic system is determined by
the output signal of another dynamic system. The new concept offers
a comprehensive explanation of physical mechanism of multi-system
interaction. The properties and potential applications of
dynamic-order fractional dynamic systems are further explored with
the analysis of anomalous relaxation and diffusion processes.
\end{abstract}

\maketitle


Fractional dynamic system has been focused by physicists and
mathematicians over the last decades, and has received great success
in the analysis of anomalous diffusion
\cite{Metzler2000,Zaslavsky2002,Magin2008,Sokolov2006}, viscoelastic
rheology \cite{Hernandez-Jimenez02,Yang10}, control systems
\cite{Chen02,Podlubny1999b}, complex networks \cite{West2008}, wave
dissipation in human tissue and electrochemical corrosion process
\cite{Kilbas06}, etc. \cite{Baleanu05,Magdziarz2008,Saichev97}.

In the past several years, various physical applications have given
birth to the variable-order fractional dynamic system
\cite{Samko1995,Lorenzo2002,Sun2009a}. However, the multi-system
interaction must be considered in the physical mechanism analysis of
the variable-order fractional dynamic system and its applications.
Especially, the behavior of a dynamic system may change with the
evolution of other dynamic systems in multi-system physical
processes. How to characterize the interaction effect between these
dynamic systems? Establishing a system of equations which includes
intricate interaction terms, will cause great difficulties in
modeling and computation. Even worse, since they may miss capturing
the critical physical mechanism of the considered problems, the
established model will produce incorrect results which greatly
deviate from experimental results or field measurement data.
Meanwhile, researchers have confirmed that the differential orders
in some fractional dynamic systems are non-constant and are often
functions of other variables or system outputs
\cite{Ingman2000,Sun2009a}. For instance, Gl$\ddot{o}$ckle and
Nonnenmacher have found that the differential order of proteins
relaxation is a function of temperature \cite{Glockle1995}.
Therefore, in order to exploit the physical mechanism of a
variable-order fractional dynamic system, another dynamic system
usually should be considered. For simplicity, we name this type of
variable-order fractional dynamic system as dynamic-order fractional
dynamic system.

The purpose of this letter is to make an innovative study of
dynamic-order fractional dynamic systems. Since previous studies
have indicated that the differential-order is a critical factor in
the variable-order fractional dynamic system, it is necessary to
analyze the physical mechanism about how environmental factors or
system variables influence the system's differential order. From the
analysis of fractional dynamic system, especially the variable-order
dynamic system, it has been found the fractional differential order
of one dynamic system usually stems from another dynamic system
\cite{Ingman2000,Sun2009a,Glockle1995}. Hence, in many fractional
systems, the differential order should be called dynamic-order. The
comprehensive study of dynamic-order system will give us new
understanding about multi-system interaction.


Firstly, we introduce the motivation of dynamic-order fractional
system via an experiment of Fractor. This experiment will give us a
basic understanding of dynamic-order fractional dynamic systems.

\begin{figure}[htb]
\centering \subfigure{
\includegraphics[width=0.45\linewidth]{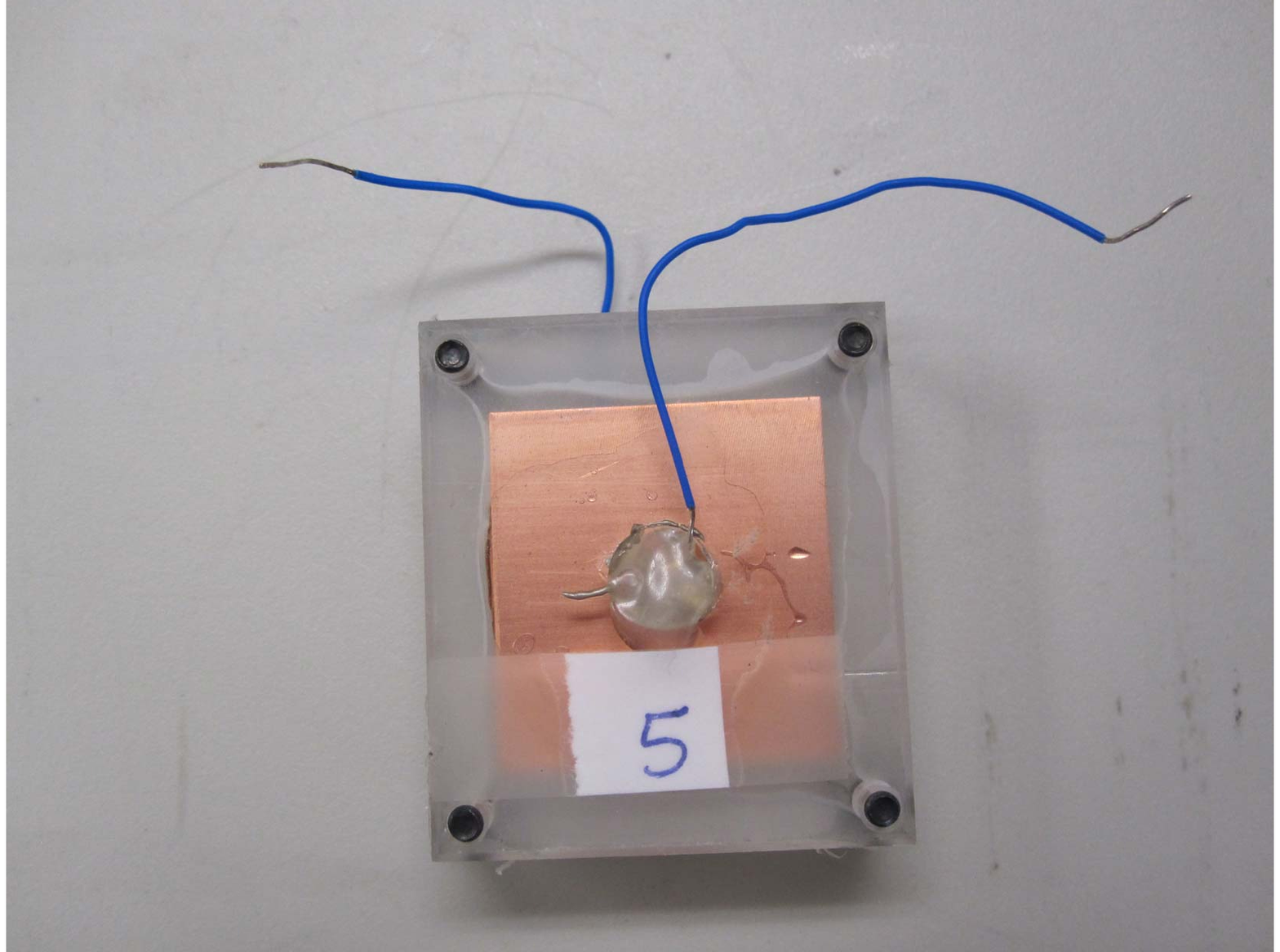}
\label{fig:forceBalanceComparison_a}} \subfigure{
\includegraphics[width=0.45\linewidth]{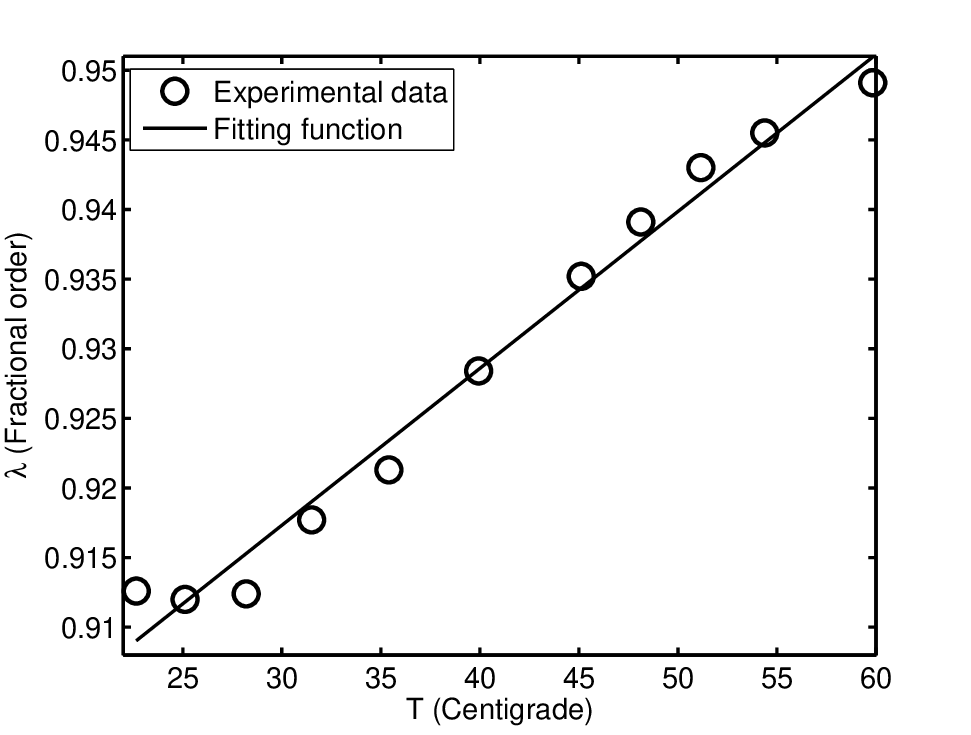}
\label{fig:forceBalanceComparison_b}} \caption{Left: Photo of a
sample hand-made prototype Fractor. The detailed preparation and
description of the Fractor can be found in \cite{Bohannan08}. Right:
The fractional order evolution data of the Fractor. In this
temperature range, the linear fitting function is $\lambda(T)=p_1
T+p_2$, where $p_1=0.001127\pm 0.000116$, $p_2=0.8835\pm 0.0048$. }
\label{fig1}
\end{figure}

A Fractor is a two lead passive electronic circuit element similar
to a resistor or capacitor, exhibiting a non-integer order power-law
impedance versus frequency \cite{Bohannan08}. The Fractor is proven
effective as a feedback element in control systems for real-world
applications such as temperature control, robotics, etc.
\cite{Sheng08}. The prototype Fractor is made by hand and not much
larger than typical through-hole capacitors and the typical unit is
$3.5$ cm on a side and about $1.0$ cm thick, as shown in Fig.
\ref{fig1} (Left). The impedance behavior of the Fractor can be
accurately modeled by the following formula which is achieved by
Laplace transform of the fractional order operator
$\displaystyle{Z_{Frac}(\omega)=K/(j \omega \tau)^\lambda }$, $K$ is
the impedance magnitude at calibration frequency
$(\omega_c=1/\tau)$; $\lambda\in (0, 1)$ is the fractional exponent
and $\omega$ is the frequency \cite{Bohannan08,Jesus2008}. From the
previous study of the Fractor, it has been confirmed that the order
of the Fractor may change with the temperature and its internal
material structure. Several experimental results have implied that,
temperature can influence the derivative order or integral order
which will determine the capacity of the equipment
\cite{Bohannan08,Glockle1995}. In our experiment, we intend to give
a clear relationship between the temperature and the fractional
order of the Fractor. In this experiment, temperature is controlled
by the equipment of Quanser Heatflow Experiment (HFE)
\cite{Bohannan08}. The evolution process of temperature is a typical
heat transfer process. In this heat transfer system, the
instantaneous temperature value of the Fractor is the output signal
we intend to obtain. Based on the previous experimental observation,
we rewrite the above formula as
\begin{eqnarray}
\begin{array}{c}
\displaystyle{Z_{Frac}(\omega)=\frac{K}{(j \omega \tau)^{\lambda(T)}
}},
\end{array}
\label{eq4}
\end{eqnarray}
where $\lambda (T)\in (0, 1)$ is the fractional exponent, $T$ is the
temperature.

In the experiment, we firstly placed the Fractor in the HFE unit,
then the temperature of the Fractor was measured by three
temperature sensors. Next, the order of the Fractor was calculated
by an HP DSA and fitted via the expression (\ref{eq4}). Finally, The
relationship between the temperature and the order of the Fractor
was obtained as shown in Fig. \ref{fig1} (Right). The experimental
details were stated in \cite{Sheng08}. It is observed that the order
of the Fractor is an approximately linear function of the
temperature. Since the order of the Fractor changes with the
temperature which is the output signal of the heat transfer system
in the Fractor, this system is a typical dynamic-order system.

Next, we discuss the dynamic-order fractional dynamic system from
the viewpoint of variable-order fractional dynamic system. A
representative definition of variable-order fractional derivative in
Caputo sense can be stated as follows \cite{Coimbra2003,Sun2010b}
\begin{eqnarray}
\displaystyle{^CD_{0+}^{\alpha(Z)}f(t)=\frac{1}{\Gamma(1-\alpha(Z))}\int_{0}^t\frac{f'(\tau){\rm
d}\tau}{(t-\tau)^{\alpha(Z)}},\,\,0<\alpha(Z)< 1,} \label{eq3}
\end{eqnarray}
where $Z$ denotes certain system variable or other independent
variable, which govern the differential-order of interested
dynamic-order system. $\alpha(Z)$ is a function of the independent
variable $Z$  and $\Gamma()$ is the Gamma function. In some
variable-order fractional dynamic systems, the variable-order is a
function of certain variable, such as temperature or concentration,
while the order-influencing variable usually can be regarded as
output signal of another dynamic system.

\begin{figure}
\centering
 \includegraphics[width=0.50\linewidth]{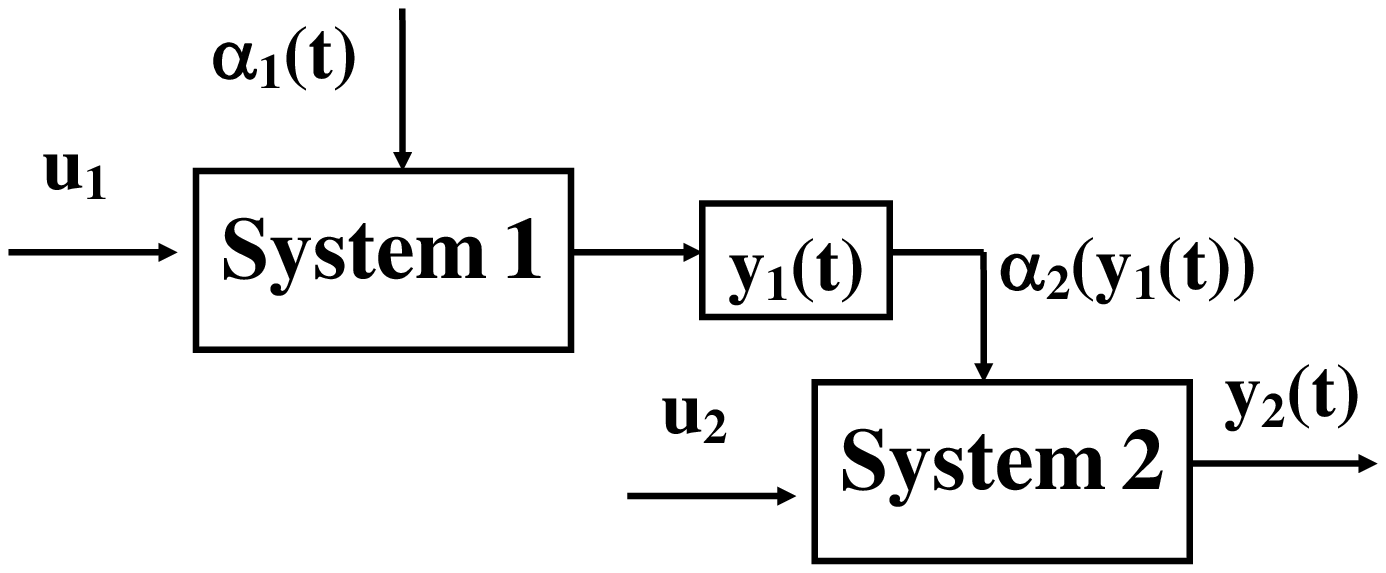}
\caption{An illustration of a dynamic-order fractional dynamic
system. In this schematic, $u_1$ is the input of system $1$ and
$\alpha_1 (t)$ is the differential order, $y_1(t)$ is the output
signal. $\alpha_2 (y_1(t))$ denotes the differential-order of system
$2$, $u_2$ and $y_2(t)$ represent the input and output signals of
system $2$.}
\label{fig4}  
\end{figure}

In engineering practice, we usually encounter the phenomena of the
multi-system interaction, such as temperature field, stress field
and electromagnetic field coupled together with the concerned
dynamic system under investigation. From the traditional approach,
when considering the coupling effect of different systems or some
multi-scale physical processes, we are accustomed to employing a
system of differential equations and they have received great
success in the past decades \cite{Meier2007}. For example, when
modeling the tracer transport in a fractured granite, Reimus et al.
\cite{Reimus03} established a convection-diffusion equation in
fracture and a diffusion equation in matrix, in which the intricate
interaction is reflected by an additional term in the
convection-diffusion equation in fracture scale. However, in the
recent decades, experimental and theoretical studies have implied
that the dynamic-order system approach can be employed to reflect
the interaction effect in some real-world multi-system problems
\cite{Bohannan08,Sun2009a}. The link between these systems may be
established via differential order of each dynamic system, which can
be visually illustrated by Fig. \ref{fig4}.

From the viewpoint of dynamic-order system, when investigating some
multi-system problems, we can establish the following generalized
form of dynamic-order fractional dynamic systems
\begin{eqnarray}
\left\{
\begin{array}{c}
\displaystyle{\frac{d^{\alpha_1 (X_{n}, X_{n-1}, ..., X_1, t)}X_1}{d
t^{\alpha_1 (X_{n}, X_{n-1}, ..., X_1, t)}}=g_1(X_1,
t),\,}\\
\displaystyle{\frac{d^{\alpha_2 (X_{n}, X_{n-1}, ..., X_1, t)}X_2}{d
t^{\alpha_2 (X_{n}, X_{n-1}, ..., X_1, t)}}=g_2(X_2,
t),\,}\\
\displaystyle{\vdots }\\
\displaystyle{\frac{d^{\alpha_n (X_{n}, X_{n-1}, ..., X_1, t)}X_n}{d
t^{\alpha_n (X_{n}, X_{n-1}, ..., X_1, t)}}=g_n(X_n, t).}\\
\end{array}\right.\;
\label{eq2}
\end{eqnarray}
The most important feature of the above generalized form is that the
intricate interaction between dynamic systems has been represented
by the differential order of each sub-system.

In the following, we further illustrate the dynamic-order fractional
dynamic system through two cases.

 \emph{Case 1.} We firstly consider
the relaxation system, which has been widely applied in energy
dissipation, viscoelasticity and rheology, etc.
\cite{Bagley83,Coimbra2003,Metzler2002}. To characterize the
relaxation process, in which the relaxation pattern changes with
control parameters or other variables, the following variable-order
fractional differential equation may be employed
\begin{eqnarray}
\left\{
\begin{array}{c}
^CD_t^{\alpha(Z)}x(t)=-B x(t)+f(t), \, 0<\alpha(Z)<1, \\
x(0)=1,
\end{array}\right.\;
\label{eq:one}
\end{eqnarray}
where $B$ is the relaxation coefficient and $Z$ is an independent
variable.

In engineering fields, fuzzy systems have extensive applications in
control and system parameter calibration, and have become an
efficient tool to exploit clear conclusion from uncertain
information and incomplete data \cite{Wang1996,Kosko98,Wang05}.
Thereby, we consider the dynamic system (\ref{eq:one}) in which the
variable-order is governed by the following fuzzy dynamic system.
Here, we select the Takagi-Sugeno (T-S) fuzzy system as an example
for illustration purpose. The T-S fuzzy system obeys the following
rule \cite{Song08}:

Plant Rule $i$: If $s_{1}(t)$ is $\mu_{i1}$ and $\cdots$ and
$s_{p}(t)$ is $\mu_{ip}$, then
\begin{equation}
\displaystyle{y(t)=A_i y_0(i), i=1,2,\ldots,N,}
\end{equation}
where $\mu_{ij}$ is the fuzzy set and $N$ is the number of IF-THEN
rules; $s_{1}(t),\ldots,s_{p}(t)$ are the premise variables. Then
the final output of T-S fuzzy system is inferred as follows:
\begin{eqnarray}
\left\{
\begin{array}{c}
\displaystyle{\dot{y}(t)=\sum_{i=1}^{N} h_i(y(t))A_i y(t),}\\
\displaystyle{y(0)=1.0.}
\end{array}\right.\;
\label{eq:four}
\end{eqnarray}
In this illustrative case, we assume $N=2$ and the fuzzy membership
$h_1(y(t))=\frac{1}{2}-\frac{1}{2}
y(t),\,h_2(y(t))=\frac{1}{2}+\frac{1}{2} y(t)$.

\begin{figure}
\centering
\includegraphics[width=0.90\linewidth]{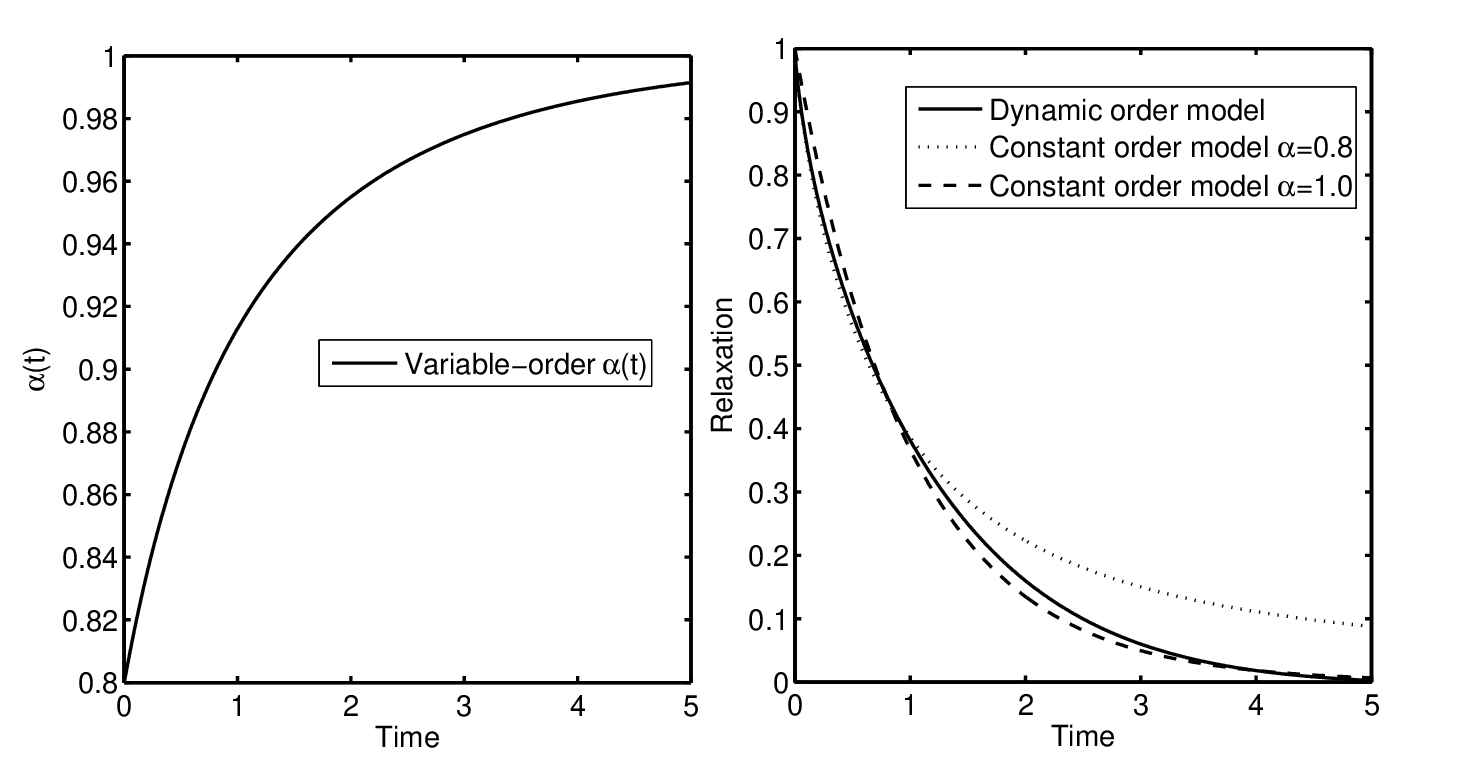}
\caption{Left: The curve of the variable-order $\alpha(t)$
 originated from the fuzzy system
(\ref{eq:four}). Right: The relaxation curve of the system
(\ref{eq:one}) with the dynamic-order $\alpha(t)$ given in
(\ref{eq6}) and $B=1.0$.}
\label{fig7} 
\end{figure}

If $A_1=0, A_2=-1$, then the exact solution of (\ref{eq:four}) is

\begin{eqnarray}
\begin{array}{c}
\displaystyle{y(t)=\frac{1}{2 e^{t/2}-1}}.
\end{array}
\label{eq13}
\end{eqnarray}
For numerical simulation, we assume that the relationship between
the output of fuzzy system (\ref{eq:four}) and variable-order of
fractional dynamic system (\ref{eq:one}) can be stated as follows
\begin{eqnarray}
\begin{array}{c}
\displaystyle{\alpha(Z)=\alpha(t)=1.0-0.2 y(t).}
\end{array}
\label{eq6}
\end{eqnarray}
The evolution curve of variable-order $\alpha(t)$ and the
corresponding numerical result of the relaxation system
(\ref{eq:one}) are shown in Fig. \ref{fig7}.

From the observation of Fig. \ref{fig7}, the system (\ref{eq:one})
exhibits accelerating relaxation behavior with the variable-order
(\ref{eq6}). Since fuzzy systems have been regarded as efficient
tools to tackle real-world engineering problems, this case implies
that, fuzzy systems may be an important method to characterize the
evolution behavior of variable-order in fractional dynamic system.

\emph{Case 2.} We further consider the variable-order fractional
diffusion system, in which the differential order is determined by a
fractional order dynamic system. It means that the differential
order in the considered fractional diffusion system relates to the
output of another fractional dynamic system. It is a more general
form to investigate the dynamic-order system. The considered
variable-order fractional diffusion system is stated as
\begin{eqnarray}
\left\{
\begin{array}{l}
\displaystyle{^CD_t^{\alpha(T)}u(x,t)=K \frac{\partial^2
u(x,t)}{\partial x^2}+q(x,t), }\\
\displaystyle{u(x,0)=\sin(x),\,
 x\in [0,L],}\\
\displaystyle{u(0,t)=u(L,t)=0,\, t\in [0,M],}
\end{array}\right.\;
\label{eq16}
\end{eqnarray}
where $K>0$ is the generalized diffusion coefficient, $u(x,t)$ is
concentration, mass or other quantities of interest, $q(x,t)$ is a
source term and $\alpha(T)$ is a function of temperature.

\begin{figure}
\centering
 \includegraphics[width=0.90\linewidth]{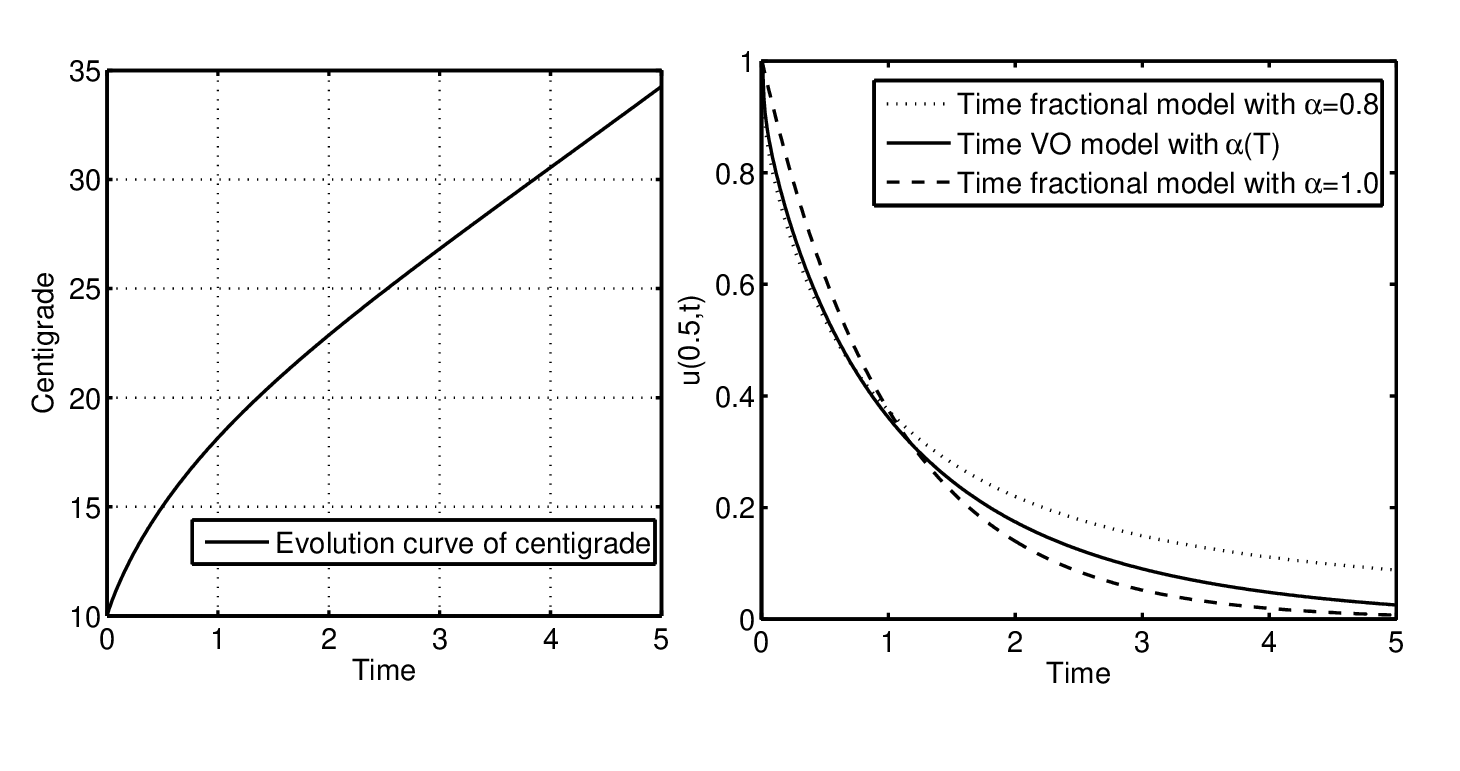}
\caption{Left: The evolution curve of the temperature (\ref{eq14})
with $\beta=0.9$. In numerical simulation, the time step is $0.01$.
Right: The diffusion curve of the system (\ref{eq16}) with
$L=1.0,\,M=5.0,\,x=0.5,\, q(x,t)=0$ and $K=0.1$, the differential
order is (\ref{eq15}).}
\label{fig9}       
\end{figure}

We suppose the evolution process of temperature can be characterized
by
\begin{eqnarray}
\left\{
\begin{array}{c}
\displaystyle{^CD_t^{\beta}T(t)=0.1 T(t)+10/(1.3 t+1)},\\
\displaystyle{T(0)=10},
\end{array}\right.\;
\label{eq14}
\end{eqnarray}
where $\beta \in (0,1]$ is the factional order derivative. Then, we
assume the relationship between the differential order (\ref{eq16})
and the temperature is
\begin{eqnarray}
\begin{array}{c}
\displaystyle{\alpha(T)=0.8+0.005 T.}
\end{array}\;
\label{eq15}
\end{eqnarray}
It indicates that the diffusion process characterized by
(\ref{eq16}) is an accelerating subdiffusion process. The trajectory
of $\alpha(t)$ and the diffusion curve of system (\ref{eq16}) at
$x=0.5$ are drawn in Fig. \ref{fig9}.

This example shows that the heat transfer process with result of
temperature increasing, causes the accelerating behavior of
considered diffusion process. In engineering situations, this model
can offer an effective tool to explore physical mechanism of
real-world diffusion related dynamic processes.

Finally, we should note that, though we have made a first attempt to
introduce the concept of dynamic-order fractional dynamic system, in
which the multi-system interaction is delivered by the differential
order of each system. The further investigations on physical
mechanism and application potentials are deserved.

H. G. Sun thank G. W. Bohannan, X. N. Song and the AFC reading group
meeting in CSOIS, Utah State University for discussions. The work is
supported by 2010CB832702, 201101014 and 2010B19114.

\bibliography{apssamp}

\end{document}